\documentclass[twocolumn,prd,aps,showpacs,showkeys,amsmath,amssymb]{revtex4-1}
\usepackage{bm}

\usepackage{color}
\usepackage{amsmath}
\usepackage{amsfonts}
\usepackage{verbatim}
\usepackage{amssymb}
\usepackage{graphicx}
\usepackage{epstopdf}
\usepackage{mathrsfs}
\usepackage{epsfig}

\providecommand{\U}[1]{\protect\rule{.1in}{.1in}}
\newcommand{\bl}{\boldsymbol}

\newcommand{\ph}{\phantom}

\newcommand{\eq}{\,=\,}
\newcommand{\ma}{\,+\,}
\newcommand{\me}{\,-\,}

\begin{document}

\title{Generalized Charged Nariai Solutions in Arbitrary Even Dimensions \\ with Multiple Magnetic Charges}
\author{Carlos Batista}
\email[]{carlosbatistas@df.ufpe.br}
\affiliation{Departamento de F\'{\i}sica, Universidade Federal de Pernambuco,
Recife 50740-560, Pernambuco, Brazil}
%\author{Andr\'{e}s Anabal\'{o}n}
%\email[]{andres.anabalon@uai.cl}
%\affiliation{Departamento de Ciencias, Facultad de Artes Liberales y Facultad de Ingenier\'{\i}a y Ciencias, Universidad Adolfo Ib\'{a}\~{n}ez,  Av. Padre %Hurtado 750, Vi\~{n}a del Mar, Chile}

%\date{\today}

\begin{abstract}
Higher-dimensional solutions for Einstein-Maxwell equations that generalize the charged Nariai spacetime are obtained. The solutions presented here are made from the direct product of several 2-spaces of constant curvature. These solutions turn out to have many magnetic charges, contrary to the usual higher-dimensional generalization of the Nariai spacetime, which has no magnetic charge at all. These solutions are then used to generate black hole metrics. Finally, it is analyzed how these generalized Nariai solutions are modified in more general theories of gravity.
\end{abstract}
\maketitle

\section{Introduction}

The use of higher-dimensional spacetimes to explain our physical world has a long history in physics. Indeed, even prior to the publication of the final form of general relativity, G. Nordstr\"{o}m made use of a five-dimensional spacetime in order to unify gravity and electromagnetism into a single scheme \cite{Nordstrom}. Few years later, under the light of general relativity, T. Kaluza and O. Klein made this possibility clearer in influential works whose concepts are of relevance until now \cite{Kaluza-Klein}. However, the idea that the universe can have more that four dimensions started to spread and be considered by a bigger community around the 70s, when string theory arose as a possibility of providing a quantum theory of gravity \cite{CremmerString}.  According to the current interpretation, the reason why we are not used to notice the six extra dimensions necessary in string theory is that they are very small and, therefore, can only be probed through highly energetic experiments. Nevertheless, from the theoretical point of view, it is also possible to reconcile our impression that we live in a four-dimensional world with the existence of large extra dimensions \cite{ArkaniHamed:1998rs,Csaki:2004ay}. In this scenario, the interactions of the standard model are restricted to a four-dimensional brane while gravity permeates all dimensions. Even infinity extra dimensions are not necessarily in contradiction with our daily experiences \cite{Randall:1999vf,Csaki:2004ay}. Nevertheless, it is fair to say that astrophysical and earth-based high energy experiments have put huge constraints in the possibility of existing extra dimensions of sizes much greater than the Planck length  \cite{Uehara:2002yv,Arun:2015ubr}. Nowadays, another important source of attention for higher-dimensional spacetimes is the AdS/CFT correspondence \cite{Maldacena:1997re}, which relates a gravitational theory in the bulk of an $n$-dimensional spacetime with a conformal field theory in the $(n-1)$-dimensional boundary. Due to all these branches of physics that make use of higher-dimensional spacetimes, it is increasing the amount of research in such subject.

%The ``problem'' is that the consistency of string theory requires the spacetime to have ten dimensions.
%This correspondence provides a great tool for performing calculations in strongly interacting conformal field theories by means of weak

The Schwarzschild-de Sitter black hole is not in thermodynamical equilibrium, since the temperatures of the black hole horizon and cosmological horizon are not the same. As the limit of equal temperatures is taken, the two horizons approach each other and the outcome of this limit process is the Nariai spacetime \cite{Ginsparg:1982rs,Hawking-Bousso:1996au}, a solution first found in Ref. \cite{Nariai}. The Nariai metric is a four-dimensional vacuum solution of Einstein's field equation in the presence of a positive cosmological constant that is the direct product of two spaces of constant curvature, namely $dS_2$ and $S^2$. Analogously, the so-called anti-Nariai, Bertotti-Robinson and Plaba\'{n}ski-Hacyan solutions are other vacuum solutions formed by the direct product of two spaces of constant curvature \cite{Bertotti-Robinson,PlebHacy,Ortaggio:2002bp}. For an account of impulsive gravitational waves in these spacetimes, see  \cite{Ortaggio:2002bp,Ortaggio:2001af}, while the thermodynamics of the Nariai spacetime have been considered in Ref. \cite{Eune:2012mv}.

An electrically charged higher-dimensional generalization of the Nariai spacetime have been obtained in Ref. \cite{Kodama:2003kk}. This solution of Einstein-Maxwell equations is the direct product of the spaces $dS_2$ and $S^{n-2}$, and can be obtained from Schwarzschild-Tangherlini black hole in the limit of equal temperatures of the horizons, as proved in \cite{Cardoso:2004uz}. Here, we will present a different higher-dimensional generalization of the Nariai solution that is formed from the direct product of $dS_2$ with several 2-spheres possessing different radii. One interesting feature of the latter solution is that, besides having an electric charge, it also admits several magnetic charges, diversely from the solutions obtained in Refs. \cite{Kodama:2003kk,Cardoso:2004uz}, which have no magnetic charge at all. In addition, we will investigate whether these new generalized Nariai spacetimes are associated to black holes and if they have counterparts in theories of gravity that are more general than Einstein's theory.

The outline of the article is the following. In Sec. \ref{Sec.NariaiEinstein}, the new higher-dimensional generalizations of Nariai solutions are presented and some of their geometrical and  physical properties are investigated. Moreover, we also obtain higher-dimensional extensions of anti-Nariai, Bertotti-Robinson and Plaba\'{n}ski-Hacyan solutions. Then, in Sec. \ref{Sec.BlackHoles}, we use the generalized Nariai metrics to obtain black hole solutions for Einstein-Maxwell equations. In Sec. \ref{Sec. f(r)-Lovelock}, we investigate metrics made from the product of 2-spaces of constant curvature that are solutions for more general gravitational theories coupled to an electromagnetic field through a minimal coupling. Finally, in Sec. \ref{Sec. Conclusioins}, we summarize the results and discuss some perspectives.

\section{Generalized Charged Nariai Solutions in Einstein's Theory} \label{Sec.NariaiEinstein}

In this starting section  we are interested in obtaining solutions for a gravitational system interacting, via minimal coupling, with an electromagnetic field and a cosmological constant $\Lambda$. The system is described by the following action
\begin{equation*}
  \mathcal{S} \eq \int \sqrt{-g} \left[  \mathcal{R} \me  (n-2) \, \Lambda \me \frac{1}{4}\,\mathcal{F}^{cd}\mathcal{F}_{cd} \right] \,,
\end{equation*}
where $\mathcal{F}_{ab} = \nabla_{[a}\mathcal{A}_{b]}$ is the electromagnetic field, $\mathcal{R}$ stands for the Ricci scalar and $n$ is the dimension of the spacetime. The field equations of this system are:
\begin{gather}
  \mathcal{R}_{ab} - \frac{1}{2}\,\mathcal{F}_a^{\ph{a}c}\mathcal{F}_{bc} =
   \frac{g_{ab}}{2}  \left[ \mathcal{R} - (n-2) \Lambda - \frac{1}{4}\,\mathcal{F}^{cd}\mathcal{F}_{cd} \right] \nonumber \\
\quad   \nonumber  \\
 \textrm{and} \quad \nabla^a\mathcal{F}_{ab} \, = \,  0 \,. \label{Eq.Motion}
\end{gather}
In order to write the solutions that will be presented in the sequel, it is useful to adopt the notation
\begin{equation*}
   d\Omega_j^2 = d\theta_j^2 \ma \sin^2\theta_j\,d\phi_j^2
\end{equation*}
to represent the line element of the unit sphere $S^2$. In what follows, we will assume that the dimension $n$ is even and that the index $j$ ranges from $2$ to $n/2$.

The first result presented here is that a static solution for the equations of motion (\ref{Eq.Motion}) is provided by the following fields:
\begin{gather}
ds^2 = R_1^2( - \sin^2x\,dt^2 + dx^2 )   +  \sum_{j=2}^{n/2}\,R_j^2\,d\Omega_j^2 \,,   \label{NariMetric}\\
\bl{\mathcal{F}} = q_1  R_1^2  \sin x  \, dt\wedge dx +   \sum_{j=2}^{n/2}\,q_j R_j^2 \sin\theta_j \,d\phi_j\wedge d\theta_j  , \nonumber
\end{gather}
where $q_1$ is an electric charge and $q_j$ are magnetic charges. The radii $R_1$ and $R_j$ are real constants that are related to the charges and the cosmological constants by the subsequent relations:
\begin{align}
 R_1 =& \left[\,  \Lambda \me \frac{1}{2}   q_1^2 \ma  \frac{Q}{2(n-2)}  \,\right]^{-1/2} \;,  \label{R1}\\
  R_j =& \left[\,  \Lambda \ma \frac{1}{2}  q_j^2 \ma  \frac{Q}{2(n-2)}  \,\right]^{-1/2}  \;, \label{Rj}
\end{align}
where
\begin{equation}\label{Q}
 Q \, \equiv\,  q_1^2 \me \sum_{j=2}^{n/2}\, q_j^2   \,.
\end{equation}
One can check that in order for the radii to be real, which assures the Lorentzian signature, the following constraint must be satisfied by the cosmological constant:
\begin{equation}\label{LambPos}
  2\,(n-2)\,\Lambda \, \geq \, (n-3) \, q_1^2 \ma \sum_{j=2}^{n/2}\,  q_j^2   \,.
\end{equation}
In particular, the cosmological constant must be positive. If $n\neq 4$, the relations (\ref{R1}) and (\ref{Rj}) can be inverted to write the charges $q_1$ and $q_j$ in terms of the radii, the final result being
\begin{align}
 q_1 =& \sqrt{ \,\, \frac{4\,(n-2)}{(n-4)} \Lambda   \me  \frac{4}{(n-4)R_0^2} \me  \frac{2}{R_1^2} \,}  \;,  \label{q1}\\
 q_j =& \sqrt{ -\,  \frac{4\,(n-2)}{(n-4)} \Lambda   \ma  \frac{4}{(n-4)R_0^2} \ma \frac{2}{R_j^2} \,} \;, \label{qj}
\end{align}
with $R_0$ being defined by
\begin{equation*}
  \frac{1}{R_0^2} \,\equiv\,  \frac{1}{R_1^2}  \ma \sum_{j=2}^{n/2}\, \frac{1}{R_j^2}  \,.
\end{equation*}
Thus, instead of considering the electromagnetic charges as being arbitrary, we can, equivalently, suppose that the radii $R_1$ and $R_j$ are arbitrary, while the charges are determined in terms of these radii by Eqs. (\ref{q1}) and  (\ref{qj}). This is quite interesting from the point of view of compactification of extra dimensions, since we can set $R_1$ and $R_2$ to be much bigger than the other radii. However, this freedom of choosing either $\{q_1,q_j\}$ or $\{R_1, R_j\}$ as being the independent parameters is not valid in the particular case $n=4$, as can already be grasped from Eqs. (\ref{q1}) and  (\ref{qj}), which diverge in such a case. Indeed, by means of Eqs. (\ref{R1}) and (\ref{Rj}) one can see that if $n=4$  the radii $R_1$ and $R_2$ are not independent from each other, they must obey the constraint
\begin{equation*}
  \frac{1}{R_1^2}   \ma  \frac{1}{R_2^2} \eq  2 \, \Lambda \,.
\end{equation*}
Thus, when $n=4$ the solution presented here reduces to the well-known charged Nariai solution \cite{Nariai}. Therefore, the metric (\ref{NariMetric}) should be seen as a higher-dimensional generalization of the Nariai solution.

Just as the Nariai solution, the higher-dimensional version of the Nariai spacetime presented here has no singularities. Indeed, its Riemann tensor and its Weyl tensor are both covariantly constant and, therefore, all invariant scalars that can be constructed from the curvature are constant. For instance, the full contractions of the products of the Riemann tensor are given by
\begin{equation}
 \mathcal{R}^{a_1b_1}_{\ph{a_1b_1}a_2b_2} \cdots \mathcal{R}^{a_nb_n}_{\ph{a_nb_n}a_1b_1}
=  2^n\,\left[ \frac{1}{R_1^{2n}} +  \sum_{j=2}^{n/2}\,\frac{1}{R_j^{2n}} \right] \label{Scalars2} \nonumber \,.
\end{equation}

These generalized Nariai spacetimes are contained in the Kundt class of metrics \cite{Book-GrifPodol-Stephani}, inasmuch as the null directions
\begin{equation*}
  \bl{\ell} =  \csc x  \,\partial_t  +    \partial_x \quad \textrm{and} \quad
  \bl{n} =     \csc x  \,\partial_t  -    \partial_x
\end{equation*}
are geodesic, shear-free, twist-free and expansion-free. They are also repeated principal null directions of the Weyl tensor \cite{Bel-Deb.Higher,Bel-Deb},
\begin{equation*}
  C_{ab[cd}\,\ell_{e]} \ell^b   = 0 = C_{ab[cd}\, n_{e]} n^b \,,
\end{equation*}
so that, according to the boost weight classification  \cite{Coley:2004jv,OrtaggioReview}, the algebraic type of the Weyl tensor is $D$. These properties are also in agreement with the four-dimensional Nariai solution.

Using the coordinates $\tau=R_1\,t$ and $r = R_1 \cos x$, the line element of the generalized Nariai spacetime is written as
\begin{equation*}
  ds^2 = -\,h(r)\,d\tau^2 + \frac{1}{h(r)} dr^2    +  \sum_{j=2}^{n/2}\,R_j^2\,d\Omega_j^2 \,,
\end{equation*}
where $h(r) = 1 - r^2/R_1^2$. In these coordinates, it is clear that the hyper-surfaces $r=\pm R_1$ are closed null surfaces and comprise event horizons. The entropies of these horizons are given by one quarter of their area,
\begin{equation}\label{Entropy1}
 S \eq \frac{1}{4}\, \prod_{j=2}^{n/2}\,4\,\pi\,R_j^2  \,.
\end{equation}
These null hyper-surfaces are Killing  horizons associated to the Killing vector $\partial_\tau$. In order to calculate the temperature of such horizons, one must compute the surface acceleration of the of this Killing vector,
\begin{equation}\label{Kappa1}
  \kappa = \left. \sqrt{-\,\frac{1}{2}\, \nabla^a \xi^b \,\nabla_a \xi_b  } \, \right|_{r=r_h} \,,
\end{equation}
where $r_h$ is the value of the coordinate $r$ at the horizon and $\xi^a$ is the Killing vector field properly normalised. A natural way to normalize $\bl{\xi} = \lambda\, \partial_\tau$ is to choose  the multiplicative constant $\lambda$ in such a way that $\xi^a\xi_a = -1$ at the value of $r=r_{\star}$ for which $\bl{\xi}$ is a geodesic vector field, just as happens at the asymptotic infinity of an asymptotically flat spacetime \cite{Hawking-Bousso:1996au}. At $r=r_{\star}$ the gravitational, electromagnetic and cosmological forces balance each other and the observer can stay still without acceleration. Computing the Christoffel symbol $\Gamma^a_{\tau \tau}$, we see that this value of $r$ is the one for which $h'(r_{\star}) = 0$, with $h'(r)$ standing for the derivative of $h(r)$ with respect to $r$. Therefore, the normalized Killing vector field and the temperature of the horizon are respectively given by
$$ \bl{\xi} = \frac{1}{\sqrt{h(r_{\star})}}\, \partial_\tau \quad \textrm{and} \quad
  T \eq \frac{\kappa}{2\pi} = \frac{|h'(r_h)|}{4\,\pi\,\sqrt{h(r_{\star})} }$$
In the case of the metric considered here, we have $r_h= \pm R_1$ and $r_{\star} = 0$, so that the temperature of the horizons of the generalized Nariai solutions considered here is
\begin{equation*}
  T = \frac{1}{2\,\pi\,R_1} \,.
\end{equation*}
This is the temperature measured by a still observer at $r=0$. Of course, other observers should measure different temperatures. For a nice account on the thermodynamics of the four-dimensional Nariai spacetime, the reader is referred to Ref. \cite{Eune:2012mv}.

We have seen that, due to Eq. (\ref{LambPos}), these generalized Nariai solutions exist only for positive cosmological constant. Nonetheless, it is also possible to make analytical continuations on the coordinates and find the analogue of the latter solution in the case of negative cosmological constant, obtaining a generalized anti-Nariai solution. Indeed, the line element (\ref{NariMetric}) is the direct product of the de Sitter space $dS_2$ with $(\frac{n}{2} -1)$ copies of the sphere $S^2$, which are both two-dimensional spaces of constant positive curvatures. Thus, in an analogous fashion, we can find a solution whose metric is the product of spaces of constant negative curvatures, namely the product of anti-de Sitter space $AdS_2$ with $(\frac{n}{2} -1)$ copies of a hyperboloid $H^2$, in which case the cosmological constant can be negative. Actually, we can go one step further and consider the \textit{ansatz} of a spacetime whose metric is the direct product of 2-spaces of arbitrary constant curvature, which provides a generalization of Bertotti-Robinson and Pleba\'{n}ski-Hacyan four-dimensional solutions \cite{Bertotti-Robinson,PlebHacy}.  In order to accomplish this goal it is useful to define the following function depending on a discrete parameter $\epsilon$:
\begin{equation}\label{Seps}
  S_{\epsilon}(\theta) = \frac{1}{\sqrt{\epsilon}}\,\sin(\sqrt{\epsilon} \,\theta) =
\left\{
                                                                                    \begin{array}{ll}
                                                                                      \sin\theta \,,\; &\textrm{if $\epsilon = 1$}  \\
                                                                                      \;\;\theta  \,\,,\;  &\textrm{if $\epsilon = 0$} \\
                                                                                      \sinh\theta  \,,\;  &\textrm{if $\epsilon = -1$}
                                                                                    \end{array}
                                                                                  \right. .
\end{equation}
Then, the line element of a constant curvature two-dimensional space can be conveniently written as
\begin{equation}\label{LineElCC}
   d\tilde{\Omega}_{\epsilon_j}^2 = d\theta_j^2 \ma S_{\epsilon_j}(\theta_j)^2\,d\phi_j^2 \,,
\end{equation}
with the case $\epsilon_j = 1$ representing a sphere, the plane being represented by $\epsilon_j = 0$, while the hyperboloid corresponds to $\epsilon_j = -1$.
Then,  we shall seek for solutions whose line elements are given by
\begin{equation}\label{NariMetric2}
 ds^2 = R_1^2\left[ - S_{\epsilon_1}(x)^2\,dt^2 + dx^2 \right]   +  \sum_{j=2}^{n/2}\,R_j^2 \, d\tilde{\Omega}_{\epsilon_j}^2 \,.
\end{equation}
One can check that metric (\ref{NariMetric2}) is a solution of the field equations (\ref{Eq.Motion}) when $n\geq6$ provided that the electromagnetic field is given by
$$  \bl{\mathcal{F}} = q_1 \, R_1^2 \,  S_{\epsilon_1}(x)  \, dt\wedge dx +   \sum_{j=2}^{n/2}\,q_j\, R_j^2\, S_{\epsilon_j}(\theta_j) \,d\phi_j\wedge d\theta_j  \,, $$
where the electric charge $q_1$ and the magnetic charges $q_j$ might be given by the following relations respectively
\begin{align}
 q_1 =& \sqrt{ \,\, \frac{4\,(n-2)}{(n-4)} \Lambda   \me  \frac{4}{(n-4)\tilde{R}_0^2} \me  \frac{2\,\epsilon_1}{R_1^2} \,}  \;,  \label{q1v2}\\
 q_j =& \sqrt{   -\,\frac{4\,(n-2)}{(n-4)} \Lambda   \ma  \frac{4}{(n-4)\tilde{R}_0^2} \ma  \frac{2\,\epsilon_j}{R_j^2}  \,} \;. \label{qjv2}
\end{align}
The constant $\tilde{R}_0$ used in the latter expression is defined by
\begin{equation}\label{R0til}
  \frac{1}{\tilde{R}_0^2} \,\equiv\,  \frac{\epsilon_1}{R_1^2}  \ma \sum_{j=2}^{n/2}\, \frac{\epsilon_j}{R_j^2}  \,.
\end{equation}
Now, let us define the following 1-forms constituting a local basis:
\begin{equation}\label{Frame}
 \left.
   \begin{array}{ll}
    \bl{e}_1 \eq R_1\, S_{\epsilon_1}(x)\, \,dt \;\;&, \quad  \tilde{\bl{e}}_{1} = R_1\, dx \\
  \bl{e}_j = R_j\, S_{\epsilon_j}(\theta_j)  d\phi_j \;\; &, \quad  \tilde{\bl{e}}_{j} = R_j\, d\theta_j \,.
   \end{array}
 \right.
\end{equation}
Then, the vector fields associated to these 1-forms by means of the metric comprise a Lorentz frame. In terms of such basis, the electromagnetic field is given by
$$ \bl{\mathcal{F}} = q_1 \,\bl{e}_1\wedge \tilde{\bl{e}}_{1} \ma \sum_{j=2}^{n/2}\,q_j\,\bl{e}_j\wedge \tilde{\bl{e}}_{j} \,.   $$
Since the components of $\bl{\mathcal{F}}$ in this Lorentz frame are constant, we interpret it as a uniform electromagnetic field throughout the spacetime.

From Eqs. (\ref{q1v2}) and (\ref{qjv2}), one can grasp that if either $\epsilon_1$ or some $\epsilon_j$ vanish then these equations cannot be inverted to write the radii $R_1$ and $R_j$ in terms of the charges  $q_1$ and $q_j$. For example, let us say that $\epsilon_1 = 0$, then the expressions of $q_1$ and $q_j$ do not depend on $R_1$, so that, generally, we cannot write $R_1$ as a function  of the electromagnetic charges. Nevertheless, if none of the 2-spaces that form the metric is flat, namely if $\epsilon_1$ and $\epsilon_j$ are all different from zero, then Eqs. (\ref{q1v2}) and (\ref{qjv2}) can be inverted and solved for the radii $R_1$ and $R_j$ in terms of the charges $q_1$ and $q_j$. In the latter case, the electromagnetic charges can be arbitrarily assigned to the solution.

A different generalization of the charged Nariai solution to higher dimensions has also been found elsewhere \cite{Kodama:2003kk,Cardoso:2004uz}, but such spacetime is the direct product of just two spaces of constant curvature, namely $AdS_2\times S^{n-2}$. On the other hand, the metric presented here is the direct product of several 2-spaces of constant curvature. It is also worth pointing out that whereas the solution presented in Refs. \cite{Kodama:2003kk,Cardoso:2004uz} admits no magnetic charge at all, which is related to the fact that the second Betti number of the sphere $S^{n-2}$ is different from zero only if $n=4$ \cite{Maeda:2010qz}, the solution obtained here has several magnetic charges. This interesting feature makes these metrics a rich arena for studying the physics of higher-dimensional spacetimes. After the release of the pre-print of this article, Marcello Ortaggio kindly warned that the purely magnetic case of the solutions presented here, namely when $q_1=0$, is contained in the broad class of solutions obtained in Ref. \cite{Brown:2013mwa}.

%%%%%%%%%%%%%%%%%%%%%%%%%%%%%%%%%%%%%%%%%%%%%%%%%%%%%%%%%%%%%%%%%%%%%%%%%%%%%%%%%%%%%%%%%%%%
%%%%%%%%%%%%%%%%%%%%%%%%%%%%%%%%%%%%%%%%%%%%%%%%%%%%%%%%%%%%%%%%%%%%%%%%%%%%%%%%%%%%%%%%%%%%
%%%%%%%%%%%%%%%%%%%%%%%%%%%%%%%%%%%%%%%%%%%%%%%%%%%%%%%%%%%%%%%%%%%%%%%%%%%%%%%%%%%%%%%%%%%%
%%%%%%%%%%%%%%%%%%%%%%%%%%%%%%%%%%%%%%%%%%%%%%%%%%%%%%%%%%%%%%%%%%%%%%%%%%%%%%%%%%%%%%%%%%%%
%%%%%%%%%%%%%%%%%%%%%%%%%%%%%%%%%%%%%%%%%%%%%%%%%%%%%%%%%%%%%%%%%%%%%%%%%%%%%%%%%%%%%%%%%%%%
%%%%%%%%%%%%%%%%%%%%%%%%%%%%%%%%%%%%%%%%%%%%%%%%%%%%%%%%%%%%%%%%%%%%%%%%%%%%%%%%%%%%%%%%%%%%
%%%%%%%%%%%%%%%%%%%%%%%%%%%%%%%%%%%%%%%%%%%%%%%%%%%%%%%%%%%%%%%%%%%%%%%%%%%%%%%%%%%%%%%%%%%%
%%%%%%%%%%%%%%%%%%%%%%%%%%%%%%%%%%%%%%%%%%%%%%%%%%%%%%%%%%%%%%%%%%%%%%%%%%%%%%%%%%%%%%%%%%%%

\section{Associated Black Hole Solutions } \label{Sec.BlackHoles}

It is well-known that the four-dimensional Nariai solution can be obtained from the Schwarzschild-de Sitter metric in the limit that the temperature of the black hole horizon coincides with the temperature of the cosmological horizon. Indeed, the Nariai solution is the metric perceived by an observer between the two horizons as they coalesce into a single hyper-surface. Since the coordinate range between the two horizons shrinks to zero as the temperatures of the horizons approach each other, a careful near horizon limit must be taken in order to obtain Nariai metric \cite{Ginsparg:1982rs,Hawking-Bousso:1996au}. Therefore, it is natural to think that a similar approach works in higher dimensions. Indeed, this was the path taken in Ref. \cite{Cardoso:2004uz} to obtain another higher-dimensional generalization of the Nariai solution. In the latter reference, the limit of equal temperatures of the horizons of the Schwarzschild-Tangherlini solution \cite{Tangherlini} have been taken and the final result was a spacetime of the form $AdS_2\times S^{n-2}$.

Here, we already have a higher-dimensional generalization of the Nariai solution. Thus, we could think the other way around: are the generalized Nariai metrics obtained in this article related to black hole solutions? In order to answer this question, let us investigate whether it is possible to attain the metric (\ref{NariMetric}) from a black hole solution with coalescing horizons. Let us start with the following general static black hole \textit{ansatz} that has the same topology of the spacetime presented in Eq. (\ref{NariMetric}):
\begin{gather}
ds^2 = -\, f(r) \, d\tilde{t}^{\,2} + \frac{1}{f(r)}\, dr^2 + \sum_{j=2}^{n/2}\,R_j(r)^2\,d\Omega_j^2 \,.  \label{MetricBH1}
\end{gather}
Then, we take the limit in which two horizons have the same temperature while the horizons coalesce into a single surface $r=r_h$.
The hallmark of a degenerate horizon is that $f(r)$ and $f'(r)$ both vanish at the horizon, so that near the horizon we can write
$$ f(r_h + \rho) \eq - \, \lambda\, \rho^2 + O(\rho^3) \,, $$
where $\lambda$ is some non-vanishing constant. Therefore, using $\rho = r - r_h$ as a coordinate while we take the near horizon limit, the general ansatz (\ref{MetricBH1}) becomes
\begin{gather}
ds^2 = \lambda\, \rho^2 \, d\tilde{t}^{\,2} - \frac{1}{\lambda\,\rho^2}\, d\rho^2 + \sum_{j=2}^{n/2}\,R_j(r_h)^2\,d\Omega_j^2 \,.  \label{MetricBH2}
\end{gather}
Then, defining $\tilde{t}= \lambda^{-1}\,e^{-t}\, \cot x$, $\rho= e^{t}\,\sin x$, $R_j= R_j(r_h)$ and $R_1= \lambda^{-1/2}$, we retrieve the generalized Nariai spacetime (\ref{NariMetric}). Therefore, we conclude that the generalized Nariai solution presented here induces the search for static black hole solutions of the form (\ref{MetricBH1}), which we do in the sequel.

%Even with the latter ansatz for the electromagnetic field, it is quite hard to integrate the field equations (\ref{Eq.Motion}) for the general case in which %the functions $R_j(r)$ are arbitrary. Nevertheless, one can check that Einstein's equation imply that the following condition holds:

Before proceeding, it is useful to introduce the following notation:
$$ \mathcal{E}_{ab} \,\equiv\, (\mathcal{R}_{ab} - \frac{1}{2}g_{ab}\mathcal{R}) - \mathcal{T}_{ab} \,,$$
where $\mathcal{T}_{ab}$ is the energy-momentum tensor of the electromagnetic field. Then, the condition $\mathcal{E}^a_{\;\;b} \eq 0$ is just Einstein's equation (\ref{Eq.Motion}).

In spite of the simplicity of the line element (\ref{MetricBH1}), it turns out that it is quite hard to integrate Einstein-Maxwell equations for this metric in the general case. Therefore, we shall make some simplifying assumptions. First, inspired by Eq. (\ref{NariMetric}), let us adopt the following \textit{ansatz} for the electromagnetic field
\begin{equation}\label{ElectroM-BH}
 \bl{\mathcal{F}} = F_1(r)  \, d\tilde{t}\wedge dr +   \sum_{j=2}^{n/2}\,F_j(r)  \sin\theta_j \,d\phi_j\wedge d\theta_j  \,.
\end{equation}
Even with the latter assumption, it is difficult to integrate Einstein's equation without making restrictions over the functions $R_j(r)$ that appear in the metric (\ref{MetricBH1}). Nevertheless, imposing equation $\mathcal{E}^{\tilde{t}}_{\;\;\tilde{t}} - \mathcal{E}^r_{\;\;r} = 0$, we find that the following condition must hold:
\begin{equation}\label{Eq.R}
  \sum_{j=2}^{n/2}\, \frac{R''_j(r)}{R_j(r)} \eq 0 \,.
\end{equation}
The simplest solution for the latter equation is attained when each term of the above sum vanishes, in which case the functions $R_j(r)$ have a linear dependence on $r$:
\begin{equation}\label{Rjab}
  R_j(r) \eq a_j\,r \ma b_j \,,
\end{equation}
where $a_j$ and $b_j$ are constants. In particular, when the constants $a_j$ are all zero we retrieve the generalized Nariai metric (\ref{NariMetric}). Another case that is not so alluring happens when just one of the constants $a_j$ is non-vanishing, in which case we obtain the direct product of the
Reissner-Nordstr\"{o}m solution with $\frac{n-4}{2}$ spheres of constant radii.

A particularly interesting case happens when we assume that $a_j$ are non-vanishing and the constants  $b_j$ in Eq. (\ref{Rjab}) are all equal. In such a case, we can redefine the origin of the coordinate $r$ so that the constants $b_j$ vanish. Then, after some calculations, the integration of the field equations leads to the following black hole solution:
\begin{gather}
ds^2 = -\, f(r) \, d\tilde{t}^{\,2} + \frac{1}{f(r)}\, dr^2 + r^2\, \sum_{j=2}^{n/2}\,d\Omega_j^2 \,,  \nonumber \\
  \bl{\mathcal{F}} = \sqrt{2(n-2)(n-3)}\, \frac{q_1}{r^{n-2}}\,d\tilde{t}\wedge dr \label{MetricBH} \\
+ 2\,q_2\,\sum_{j=2}^{n/2}\,\sin\theta_j \,d\phi_j\wedge d\theta_j \nonumber \,,
\end{gather}
where the constants $q_1$ and $q_2$ are arbitrary parameters proportional to the electric and magnetic charges respectively, while the function $f(r)$ is given by
\begin{equation}
  f(r) = \frac{1}{n-3} - \frac{2\,\mu}{r^{n-3}} + \frac{q_1^2}{r^{2(n-3)}}-\frac{q_2^2}{(n-5)r^2} - \frac{\Lambda\,r^2}{n-1}  \,, \label{f}
\end{equation}
with $\mu$ being an arbitrary constant. The latter parameter is proportional to the mass, i.e., it is proportional to the conserved charge associated to the Killing vector field $\partial_{\tilde{t}}$. One interesting feature of this solution is that it only has one magnetic charge $q_2$, whereas in the generalized Nariai solution (\ref{NariMetric}) a different charge can be assigned for each sphere factor in the line element. As a consequence, all sphere factors in the black hole solution (\ref{MetricBH}) must have the same radius. Note that we started the integration process with arbitrary coefficients $a_j$ and, in principle, they were independent from each other. Nonetheless, the field equations impose that all the constants $a_j$ must coincide, leading to the same radius and magnetic charge for each sphere factor. This black hole solution has already been obtained in Ref. \cite{Ortaggio:2007hs}, as a particular case of a broader solution in which the $\left(n-2\right)$-dimensional spatial part of the metric is an Einstein manifold possessing an almost-K\"{a}hler structure. Moreover, the cases $q_2=0$ (purely electric) and $q_1=0$ (purely magnetic) of this black hole solution can be attained in the limit of Einstein-Maxwell theory from the solutions found in Ref. \cite{Maeda:2010qz}.

The spacetime presented in Eq. (\ref{MetricBH}) has a singularity at $r=0$, which is indicated by the fact that the Kretschmann scalar diverges in the limit $r\rightarrow 0$. Also, the null vector fields
\begin{equation}\label{lnBH}
  \bl{\ell} =  \frac{1}{\sqrt{f}} \,\partial_{\tilde{t}}  +    \sqrt{f} \,\partial_r \quad \textrm{and} \quad
 \bl{n} =  \frac{1}{\sqrt{f}} \,\partial_{\tilde{t}}  -    \sqrt{f} \,\partial_r
\end{equation}
are repeated principal null directions of the Weyl tensor \cite{Bel-Deb.Higher,Bel-Deb},
\begin{equation*}
  C_{ab[cd}\,\ell_{e]} \ell^b   = 0 = C_{ab[cd}\, n_{e]} n^b \,.
\end{equation*}
Therefore, according to the boost weight classification \cite{Coley:2004jv,OrtaggioReview}, the algebraic type of this spacetime is $D$. The latter classification is a generalization of the Petrov classification to arbitrary dimension. These repeated principal null directions are geodesic, shear-free and twist-free, but have non-zero expansion. Thus, the solution presented here is contained in the Robinson-Trautman class of spacetimes \cite{RobinsonTrautman,RobinsonTrautman-HighD}.

In the special case $n=4$, the solution (\ref{MetricBH}) reduces to the well-known Reissner-Nordstr\"{o}m solution in the presence of a cosmological constant, with $\mu$ denoting the mass, $q_1$ is the electric charge and $q_2$ is the magnetic charge. However, for $n>4$  the solution presented here differs, in several respects, from the charged version of the Schwarzschild-Tangherlini spacetime. First, the topology of the spatial infinity and of the horizons are different in both cases. For instance, while the topology of the horizon in Schwarzschild-Tangherlini spacetime is $\mathbb{R}\times S^{n-2}$, in the solution presented here the topology of the horizon is the cartesian product of a real line with several 2-spheres. Second, note that the additive constant term in the function $f(r)$ is different from $1$ if $n\neq4$, and this constant factor cannot be modified by a coordinate transformation. Moreover, a higher-dimensional Schwarzschild metric possessing a magnetic charge does not exist, which is related to the fact that the second Betti number of the sphere $S^{n-2}$ is different from zero only if $n=4$ \cite{Maeda:2010qz}. Regarding the magnetic charge $q_2$, it is also worth stressing that the sign of in front of $q_2^2$ in Eq. (\ref{f}) becomes negative if $n\geq 6$, which implies the existence of an event horizon even in the case of vanishing mass parameter $\mu$.

Besides the Killing vector field $\partial_{\tilde{t}}$, each sphere factor of the metric (\ref{MetricBH}) provides three space-like Killing vector fields for the spacetime, so that the total number of Killing vector fields is $\frac{3}{2}n-2$. On the other hand, the Schwarzschild-Tangherlini spacetime possesses $\frac{1}{2}n^2-\frac{3}{2}n+2$ Killing vectors. Besides, for each sphere factor of the line element (\ref{MetricBH}) there exists a Killing-Yano tensor given by
\begin{equation*}
  \bl{Y}_j \eq r^3\,\sin\theta_j\,d\theta_j\wedge d\phi_j \,.
\end{equation*}
Nevertheless, the conserved quantities along the geodesic motion associated to these Killing-Yano tensors are not independent from  the ones generated by the Killing vector fields.

%%%%%%%%%%%%%%
%%%%%%%%%%%%%%
%%%%%%%%%%%%%%
%%%%%%%%%%%%%%
%%%NEW%%%%%%%%%%%%%%%%%%%%%%%%%%%%%%%%%%%%%%%%%%%%%%%%%%%%%%%%%%%%%%%%%%%%%%

Finally, let us analyse some thermodynamic properties of the black hole solution (\ref{MetricBH}). In four dimensions, its is simple matter to interpret a magnetic charge and its associated potential, due to the electric/magnetic duality. Nevertheless, in higher dimensions these two phenomena have different features, and the physical interpretation of the thermodynamic variable conjugated to a magnetic charge is trickier. Hence, for simplicity, in what follows we will consider that the magnetic charge is zero, $q_2 =0$. Being $r_h$ the radius of the horizon, namely, the value of the larger root of $f(r)$, the entropy of the horizon is just one quarter of its area:
\begin{equation*}\label{EntropyBH}
 S \eq \frac{1}{4}\, \prod_{j=2}^{n/2}\,4\,\pi\,(r_h)^2  \,.
\end{equation*}
In its turn, the temperature is given by $T\eq\kappa/(2\pi)$, where $\kappa$ is the surface acceleration at the horizon, given by the expression (\ref{Kappa1}) with the Killing vector being $\bl{\xi} = \partial_{\tilde{t}}$. Since the coordinate $r$ is not suitable at the horizon, we must use another coordinate system in order to do the calculation of the temperature properly. For instance, using Eddington-Finkelstein coordinates we easily find
$$  T \eq \frac{1}{2}\, f'(r_h) \,. $$
In the presence of electric charges, the electromagnetic field obeys $ d \star \bl{\mathcal{F}} = 16\pi \star\bl{\mathcal{J}}$, where $\bl{\mathcal{J}}$ is the 1-form representing the electromagnetic current. Then, if $\Omega$ is a space-like slice of the spacetime, the electric charge is given by:
\begin{align*}
 Q & \eq \int_{\Omega} \star\bl{\mathcal{J}} \eq
\frac{1}{16\pi} \int_{\partial\Omega}  \star \bl{\mathcal{F}} \\
& \eq \frac{(4\pi)^{\frac{n-2}{2}}}{16\pi}\,\sqrt{2(n-2)(n-3)}\,\,q_1 \,.
\end{align*}
Denoting the 1-form potential of the electromagnetic field by $\bl{\mathcal{A}}$, the conjugated thermodynamic variable of the electric charge is the electric potential:
$$  \Phi \eq \xi^a\,\mathcal{A}_a|_{r=r_h} \me  \xi^a\,\mathcal{A}_a|_{r=\infty}  \eq \sqrt{\frac{2(n-2)}{n-3}}\,\frac{q_1}{(r_h)^{n-3}} \,. $$
Finally, the Komar mass of the spacetime is given by:
$$ M \eq \frac{n-2}{8\pi\,(n-3)} \, \int dS_{ab}\, \nabla^a\,\xi^b \eq \frac{n-2}{8\pi}\,(4\pi)^{\frac{n-2}{2}}\,  \mu \,, $$
where a divergent piece arising from the cosmological constant has been subtracted, in accordance with the usual procedure \cite{Kastor:2008xb}.
Using these expressions, we can check that the first law of thermodynamics holds, namely
$$  dM \eq T\,dS \ma \Phi\,dQ \,. $$
Concerning the Smarr formula, one would expect that a relation of the type $M = \alpha_1 T S \ma \alpha_2 \Phi Q$ would be valid, for some constants $\alpha_1$ and $\alpha_2$. Nevertheless, this is no true. The reason is that the cosmological constant $\Lambda$ should also be considered a thermodynamical variable, which is identified as the analogous of pressure \cite{Dolan:2012jh,KubizVolume}. Indeed, the pressure is generally defined by
$$  P \eq  -\, \frac{\Lambda}{8\pi}\,.   $$
In this formalism, the mass is identified with the enthalpy, rather than the energy, so that the variable thermodynamically conjugated to $P$, the ``volume'',  is defined by
\begin{equation}\label{VolumeTh}
 V \eq \left( \frac{\partial M}{\partial P} \right)_{S,Q} \eq
\frac{n-2}{2(n-1)} \, (4\pi)^{\frac{n-2}{2}}\, (r_h)^{n-1} \,.
\end{equation}
In the latter derivation, its has been used the relation $f(r_h)=0$ to write $M$ in terms of $S$, $Q$ and $P$. With these definitions, one can check that
the broader first law
$$ dM \eq T\,dS \ma \Phi\,dQ  \ma V \, dP  $$
holds, as well as the Smarr formula
$$  M \eq \frac{n-2}{n-3} \, S\, T \ma  \Phi\, Q \me \frac{2}{n-3}\, P\,V \,,$$
which is in perfect accordance with the general results of Ref. \cite{KubizVolume}. As a final remark on the thermodynamical aspects of the solution (\ref{MetricBH}), note that the thermodynamical volume (\ref{VolumeTh}) is generally different from the geometric volume of the space inside the horizon, which is
$$ \mathcal{V}_{\textrm{Geom}} \eq \int_0^{r_h}  dr  \prod_{j=2}^{n/2}\,r^2\,d\Omega_j^2
\eq  \frac{2}{n-2} \, V\,. $$
Thus, the coincidence between the thermodynamical volume and the geometric volume occurs just in four dimensions, whereas in higher dimensions $\mathcal{V}_{\textrm{Geom}}$  is smaller than $V$. The discrepancy between these volumes should not raise any concern at all. For instance, it is known that for rotating black holes this difference also occurs \cite{Dolan:2012jh}. The nice thing of the solution presented here is that it is a static black hole in which the geometric volume and the thermodynamical volume do not agree, differently from the Schwarzschild-Tangherlini spacetime.

%%%FinishNew%%%%%%%%%%%%%%%%%%%%%%%%%%%%%%%%%%%%%%%%%%%%%%%%%%%%%%%%%%%%%%%%%%%%%%%%%%%%%%%%%%%%
%%%%%%%%%%%%%%
%%%%%%%%%%%%%%
%%%%%%%%%%%%%%
%%%%%%%%%%%%%%

In order to obtain solution (\ref{MetricBH}), we have assumed that the constants $b_j$ defined by Eq. (\ref{Rjab}) are all equal. Nevertheless, it is worth pointing out that there are also solutions for the general case in which the constants are distinct. As an example, it will be shown a solution in six dimensions, $n=6$. Assuming that $a_2=a_3=a$, where $a$ is non-zero, it follows that we can always redefine the origin of the coordinate $r$ in such a way that $b_3 = - b_2 = b$, so that
\begin{equation}\label{R2aa}
 R_2(r) \eq a\,r \me b \quad   \textrm{and}    \quad R_3(r) \eq a\,r \ma  b   \,.
\end{equation}
In such a case, one can check that a solution is provided by the ansatz (\ref{MetricBH1}) along with the electromagnetic field (\ref{ElectroM-BH}), where the functions $F_1$, $F_j$ and $f$ are fixed by the integration process. Indeed, imposing Maxwell's equation (\ref{Eq.Motion}) we find that
\begin{equation}\label{F1-6D}
   F_1(r) \eq  \sqrt{\frac{8}{3}}\, \frac{q_1}{(b^2 \me a^2 \,r^2)^2} \,,
\end{equation}
where $q_1$ is an integration constant that represents the electric charge. Then, imposing relations $\mathcal{E}^t_{\;\;t} + \mathcal{E}^{\theta_2}_{\;\;\theta_2} = 0$, $\mathcal{E}^t_{\;\;t} + \mathcal{E}^{\theta_3}_{\;\;\theta_3} = 0$ and $\mathcal{E}^{\theta_2}_{\;\;\theta_2} + \mathcal{E}^{\theta_3}_{\;\;\theta_3} = 0$, which are immediate consequences of Einstein's equation, we end up with the following relations respectively
\begin{widetext}
\begin{align}
 F_2(r) \eq &  \frac{b-a r}{b + a r} \bigg[ -8 \Lambda  \left(b^2-a^2 r^2\right)^2 + 6 a^2 r^2 + 4 a b
   r + 6 b^2  - 2 a^2 \left(9 a^2 r^2+2 a b r-3 b^2\right) f  \nonumber \\
 & \phantom{1} \quad\quad\quad   + 2 a (b^2-a^2 r^2) (5 a r+b) f'  -\left(b^2-a^2 r^2\right)^2 f''   \bigg]^{1/2} \,,\nonumber \\
 \nonumber \\
F_3(r) \eq &  \frac{b+a r}{b - a r} \bigg[  -8 \Lambda  \left(b^2-a^2 r^2\right)^2 + 6 a^2 r^2 - 4 a b
   r + 6 b^2  + 2 a^2  \left(-9 a^2 r^2+2 a b r+3 b^2\right)f   \label{Solution2BH}  \\
   & \phantom{1} \quad\quad\quad  + 2 a  (b^2-a^2 r^2) (5 a r - b) f' -\left(b^2-a^2 r^2\right)^2 f'' \bigg]^{1/2} \,,\nonumber \\
 \nonumber \\
f(r) \eq & \frac{1}{15 \left(b^2-a^2 r^2\right)} \,\bigg\{ C_1 -\frac{1}{2} C_2 \log \left(\frac{a r+b}{b-ar} \right)
+  \frac{4 b^2}{a^2} \left(4 b^2 \Lambda -5\right) \log \left(b^2-a^2
   r^2\right)+\frac{5 q_1^2}{4 a^2 b^4}  \log ^2\left(\frac{a r+b}{b-a r}\right) \nonumber \\
& \quad - \frac{1}{a^2 b^4 \left(b^2-a^2 r^2\right)} \bigg[ 14 a^2 b^8 \Lambda  r^2+a^2 b^6 r^2 \left(5-17 a^2 \Lambda
   r^2\right)
  -5 b^2 q_1^2 + a^4 b^4 r^4 \left(3 a^2 \Lambda  r^2-5\right)  \bigg] \bigg\} \nonumber
\end{align}
\end{widetext}
with $C_1$ and $C_2$ being integration constants. The remaining components of Einstein's equation are immediately satisfied once Eqs. (\ref{F1-6D}) and (\ref{Solution2BH}) are assumed to hold. Computing the Ricci scalar and the Kretschmann scalar we see that they diverge at $r \eq \pm b/a$, indicating that this spacetime has two singularities. Besides the latter solution, it has been checked that solutions also exist in six and eight dimensions for the general case in which  the constants $a_j$ and $b_j$ are different from each other, contrasting with the assumption made in Eq. (\ref{R2aa}). Nevertheless, the expressions for the functions $F_j$ and $f$ turn out to be quite messy in such a broad case, so that they will not be presented here.

%%%%%%%%%%%%%%%%%%%%%%%%%%%%%%%%%%%%%%%%%%%%%%%%%%%%%%%%%%%%%%
%%%%%%%%%%%%%%%%%%%%%%%%%%%%%%%%%%%%%%%%%%%%%%%%%%%%%%%%%%%%%%
%%%%%%%%%%%%%%%%%%%%%%%%%%%%%%%%%%%%%%%%%%%%%%%%%%%%%%%%%%%%%%
%%%%%%%%%%%%%%%%%%%%%%%%%%%%%%%%%%%%%%%%%%%%%%%%%%%%%%%%%%%%%%
%%%%%%%%%%%%%%%%%%%%%%%%%%%%%%%%%%%%%%%%%%%%%%%%%%%%%%%%%%%%%%
%%%%%%%%%%%%%%%%%%%%%%%%%%%%%%%%%%%%%%%%%%%%%%%%%%%%%%%%%%%%%%

\section{Generalized Nariai Spacetimes in Higher Order Curvature Theories } \label{Sec. f(r)-Lovelock}

It is widely believed that Einstein's theory of gravitation is just the low-energy limit of a more complete (quantum) theory that is valid up to energies of the order of the Planck scale, as exemplified by String theory \cite{BeckerBook}. In this scheme, the Einstein-Hilbert Lagrangian density is supplemented by terms of higher order in the curvature. Since these modified theories of gravitation have equations of motion that differ from Einstein's equation, it turns out that, generally,  vacuum solutions of the latter are not vacuum solutions of the former theories. Nevertheless, there are special metrics called \emph{universal} that, apart from a possible rescalement by a constant multiplicative factor, are vacuum solutions in any gravitational theory arising from an action that is invariant under diffeomorphisms \cite{Coley-Universal}.  For instance, the maximally symmetric spacetimes, and the Kerr-Schild metrics that are in the Kundt class are examples of universal solutions \cite{Gurses:2016moi}. These universal spacetimes are of great physical interest because they can be used as consistent vacuum states for the quantum theory of gravity, irrespective of its form. Given the relevance of these issues, in this section we shall investigate the generalized charged Nariai solution in a much broader theory of gravity than general relativity, and we will show that the uncharged generalized (anti-)Nariai solution presented here is a universal metric.

The Einstein-Hilbert Lagrangian density is just a linear function of the Ricci scalar $\mathcal{R}$. The simplest way to modify it is to consider that the gravitational Lagrangian density is a more general function of the Ricci scalar, $f(\mathcal{R})$, where $f$ could be chosen on phenomenological and experimental grounds. So, the action of the system in such a case is given by
$$ \mathcal{S} \eq \int \sqrt{-g} \left[ \, f(\mathcal{R}) \ma \mathcal{L}_{\textrm{matter}}\, \right] \,,  $$
where $\mathcal{L}_{\textrm{matter}}$ is the Lagrangian density of the matter content. Performing the variation of this action with respect to the metric, we end up with the following field equation
\begin{align}\label{Eq.motion-f(R)}
 f'(\mathcal{R}) \, \mathcal{R}_{ab}  - \left[ \nabla_a\,\nabla_b - g_{ab}\,\nabla_c\nabla^c \right]f'(\mathcal{R}) & \nonumber \\
- \frac{1}{2}\,f(\mathcal{R})\,g_{ab} & \eq \mathcal{T}_{ab}\,,
\end{align}
with $\mathcal{T}_{ab}$ denoting the energy-momentum tensor of the matter, which is obtained from the following relation:
$$ \mathcal{T}_{ab} \eq \frac{-1}{\sqrt{-g}}\, \frac{ \delta \mathcal{S}_{\textrm{matter}}}{\delta g^{ab}}  \,.  $$
It is worth pointing out that the field equation (\ref{Eq.motion-f(R)}) generally involves derivatives of metric that are higher than second order. Thus, it is natural to imagine that this feature would lead to a theory whose initial value problem is not well-posed. However, it can be proved that $f(\mathcal{R})$ theory is equivalent to Einstein's gravity coupled non-minimally with a scalar field \cite{Sotiriou:2008rp}, so that the Cauchy problem is well established, at least in four dimensions \cite{Salgado:2005hx}.

Another interesting way to systematically add higher order curvature terms to the Einstein-Hilbert Lagrangian is provided by the so-called Lovelock gravity \cite{Lovelock:1971yv}, whose gravitational Lagrangian density is given by
\begin{equation*}\label{LovelockLag}
   \mathcal{L}_{\textrm{Lov.}} \eq \sum_{k=0}^{\frac{n-2}{2}} \, \frac{\alpha_k}{2^k} \,\delta^{c_1d_1\cdots c_k d_k}_{a_1b_1\cdots a_k b_k}\,
\mathcal{R}^{a_1b_1}_{\ph{a_1a_1}c_1d_1} \cdots \mathcal{R}^{a_{k}b_{k}}_{\ph{a_1a_2}c_{k}d_{k}}  \,,
\end{equation*}
where $\alpha_k$ are constants that should be fixed by phenomenology. In $n$ dimensions, this sum could continue up to values $k\geq n/2$. However, these extra terms would not contribute to the field equations, since they are purely topological. The great feature that defines Lovelock gravity is that the above Lagrangian is the most general one that provides a field equation that is of second  order on the derivatives of the metric \cite{Lovelock:1971yv}. The term $k=0$ in the above sum represents the cosmological constant part of the Lagrangian, while the term $k=1$ gives the Ricci scalar. Thus, the term $k=2$ is the first ``non-conventional'' term and is called the Gauss-Bonnet (GB) Lagrangian density, whose expression is given by:
$$   \mathcal{L}_{GB}\eq \left( \, \mathcal{R}^2   - 4 \,\mathcal{R}^{ab}\mathcal{R}_{ab} + \mathcal{R}^{abcd}\mathcal{R}_{abcd}\, \right) \,.  $$
%\delta^{c_1d_1  c_2 d_2}_{a_1b_1  a_2 b_2}\, R^{a_1b_1}_{\ph{a_1a_1}c_1d_1}  R^{a_{2}b_{2}}_{\ph{a_1a_2}c_{2}d_{2}}  \eq
In four dimensions, this term is purely topological and, hence, does not contribute to the gravitational field equation. Nevertheless, in higher dimensions it
changes Einstein's field equation. It has recently been proved that Lovelock gravity can be seen as general relativity coupled with skew-symmetric auxiliary fields \cite{Brustein:2012uu}, but these form fields are non-dynamical \cite{DiCasola:2013yga}.

In order to cover a great amount of gravitational theories, here we shall consider that the gravitational Lagrangian density is given by a sum of the term $f(\mathcal{R})$ with the Gauss-Bonnet term. Considering that the gravitational field is interacting with an electromagnetic field through a minimal coupling, we end up with the following final action
\begin{equation}\label{Action-fGB}
  \mathcal{S} \eq \int \sqrt{-g} \left[  f(\mathcal{R}) \ma \alpha\, \mathcal{L}_{GB}  \me \frac{1}{4}\,\mathcal{F}^{cd}\mathcal{F}_{cd} \right] \,,
\end{equation}
where $f$ is an arbitrary function and $\alpha$ is some arbitrary constant. The field equations in such a case are given by $\nabla^a \mathcal{F}_{ab}\eq 0$ along with:
\begin{widetext}
\begin{equation}\label{Eq.motion-LVf}
  f'(\mathcal{R}) \, \mathcal{R}_{ab}  - \left[ \nabla_a\,\nabla_b - g_{ab}\,\nabla_c\nabla^c \right]f'(\mathcal{R})
- \frac{1}{2}\,f(\mathcal{R})\,g_{ab} \ma \alpha \, \mathcal{H}_{ab}  \eq \mathcal{T}_{ab}\,,
\end{equation}
where the tensor $\mathcal{H}_{ab}$ is defined by
\begin{equation*}
  \mathcal{H}_{ab} \eq 2 \left( \mathcal{R} \mathcal{R}_{ab} - 2 \mathcal{R}_{ac}\mathcal{R}^{c}_{\ph{c}b} - 2
    \mathcal{R}_{acbd}\mathcal{R}^{cd} +  \mathcal{R}_{acde}\mathcal{R}_{b}^{\ph{b}cde} \right) \me \frac{1}{2}\,g_{ab}\, \mathcal{L}_{GB}\,,
\end{equation*}
and the electromagnetic energy-momentum tensor is given by
\begin{equation*}
  \mathcal{T}_{ab} \eq \frac{1}{2}\,\mathcal{F}_a^{\ph{a}c}\mathcal{F}_{bc} - \frac{1}{8}\,g_{ab}\,\mathcal{F}^{cd}\mathcal{F}_{cd} \,.
\end{equation*}
We shall integrate these field equations starting with the following ansatz for the metric and for the electromagnetic field respectively:
\begin{equation*}
 ds^2 = R_1^2\left[ - S_{\epsilon_1}(x)^2\,dt^2 + dx^2 \right]   +  \sum_{j=2}^{n/2}\,R_j^2 \, d\tilde{\Omega}_{\epsilon_j}^2 \;\;, \quad
 \bl{\mathcal{F}} = q_1 R_1^2 \,  S_{\epsilon_1}(x)  \, dt\wedge dx + \sum_{j=2}^{n/2} q_j R_j^2\,  S_{\epsilon_j}(\theta_j)\,  d\phi_j\wedge d\theta_j ,
\end{equation*}
%\begin{equation*}
% \bl{\mathcal{F}} = q_1 R_1^2   S_{\epsilon_1}(x)   dt\wedge dx + \sum_{j=2}^{n/2} q_j R_j^2  S_{\epsilon_j}(\theta_j)  d\phi_j\wedge d\theta_j ,
%\end{equation*}
where $R_1$ and $R_j$ are constants, the functions $S_{\epsilon}$ have been defined in Eq. (\ref{Seps}) and the line elements $d\tilde{\Omega}_{\epsilon}^2$ are the ones presented in Eq. (\ref{LineElCC}).
Plugging such fields into the equation of motion for the electromagnetic field, $\nabla^a \mathcal{F}_{ab}\eq 0$, and into the field equation for the gravitational field, Eq. (\ref{Eq.motion-LVf}), lead us to the conclusion that these fields are a solution to such equations of motion provided that the electric charge $q_1$ and the magnetic charges $q_j$ are respectively given by
$$ q_1 =  \sqrt{- \frac{4}{n-4} \,  f( {\scriptstyle 2/\tilde{R}_0^2})
+ \left[ \frac{4\, }{(n-4)\tilde{R}_0^2 } - \frac{2 \epsilon_1}{R_1^2}  \right]\, f'( {\scriptstyle 2/\tilde{R}_0^2}) +
\left( \frac{1 }{\tilde{R}_0^2 } - \frac{\epsilon_1}{R_1^2}  \right) \frac{8\,\alpha\,\epsilon_1 }{R_1^2} } \,, $$
$$  q_j =  \sqrt{ \, \frac{4}{n-4} \,  f( {\scriptstyle 2/\tilde{R}_0^2})
- \left[ \frac{4\, }{(n-4)\tilde{R}_0^2 } - \frac{2 \epsilon_j}{R_j^2}  \right]\, f'( {\scriptstyle 2/\tilde{R}_0^2}) -
\left( \frac{1 }{\tilde{R}_0^2 } - \frac{\epsilon_j}{R_j^2}  \right) \frac{8\,\alpha\,\epsilon_j }{R_j^2} } \,,$$
\end{widetext}
where $\tilde{R}_0$ was defined in Eq. (\ref{R0til}). In particular, note that these expressions reduce to the ones presented in Eqs. (\ref{q1v2}) and (\ref{qjv2}) when $\alpha$ vanishes and $f(\mathcal{R}) = \mathcal{R} - (n-2)\Lambda$.  At this point, it is pertinent to mention that the entropy of horizons in theories of gravity with Lagrangian densities possessing higher order curvature terms is not just one quarter of the area \cite{Jacobson-Entropy,Wald-Entropy},  so that Eq. (\ref{Entropy1}) is not valid in the context of this section. Indeed, the total entropy will be the sum of a term arising from the Gauss-Bonnet part of the Lagrangian, which is essentially the integral of the Ricci scalar along the horizon \cite{Jacobson-Entropy}, plus a term arising from the $f(R)$ part of the Lagrangian, which is essentially the integral of $f'(R)$ along the Horizon \cite{MohseniSadjadi:2007zq}. Since the expression for the entropy is not particularly illuminating, we will omit it here.

In the particular case of vanishing electromagnetic field, in which $q_1=0$ and $q_j=0$, the gravitational field equation imposes that the radii  $R_1$ and $R_j$ must be all equal to each other as well as the parameters $\epsilon_1$ and $\epsilon_j$ must all coincide. In such a case, the only free parameters are $R_1$ and $\epsilon_1$, and the metric is given by
\begin{align}
  ds^2 = R_1^2\bigg[ -  S_{\epsilon_1}&(x)^2\, dt^2 + dx^2   \nonumber \\
& +  \sum_{j=2}^{n/2}  \left( d\theta_j^2 + S_{\epsilon_1}(\theta_j)^2 d\phi_j^2 \right) \bigg] \label{Nariai3}
\end{align}
Irrespective of the parameters $R_1$ and $\epsilon_1$, it turns out that the above metric is such that the left hand side of the field equation (\ref{Eq.motion-LVf}) is equal to the metric times a constant  multiplicative factor. This means that, given some  $R_1$ and  $\epsilon_1$, we can always choose the value of the cosmological constant in such a way that the field equation is satisfied. Actually, the same holds for any gravitational theory, not only for the ones covered by the action (\ref{Action-fGB}). Indeed, since the spacetime (\ref{Nariai3}) is the direct product of $n/2$  metrics of two-dimensional maximally symmetric spaces possessing the same Ricci scalar, it is a universal spacetime, as recently proved in Ref. \cite{Hervik:2015mja}. The proof goes as follows. The Riemann tensor of this spacetime is the ``sum'' of the Riemann tensors associated to each these 2-spaces. More precisely, the Riemann tensor is given by:
$$ \mathcal{R}_{abcd} = \mathcal{R}_{\bl{1}\,abcd} +  \sum_{j=2}^{n/2} \mathcal{R}_{\bl{j}\,abcd} \,, $$
where
\begin{equation*}
  \left.
     \begin{array}{ll}
      \mathcal{R}^{\ph{\bl{1}}ab}_{\bl{1}\ph{ab}cd}  \eq
         - 4\,\epsilon_1 R_1^{-2} \,\, \bl{e}_1^{\;[a}\tilde{\bl{e}}_1^{\;b]}\bl{e}_{1\,[c}\tilde{\bl{e}}_{1\,d]}\,, \\
\\
     \mathcal{R}^{\ph{\bl{j}}ab}_{\bl{j}\ph{ab}cd}  \eq
          4\,\epsilon_1 R_1^{-2} \,\, \bl{e}_j^{\;[a}\tilde{\bl{e}}_j^{\;b]}\bl{e}_{j\,[c}\tilde{\bl{e}}_{j\,d]}\,,
     \end{array}
   \right.
\end{equation*}
where the Lorentz frame (\ref{Frame}) has been used. Thus, any symmetric tensor of rank two constructed from contractions of the curvature tensor is a sum of the analog contractions on the Riemann tensors of the 2-spaces. For example,
$$ \mathcal{R}_{acde}\mathcal{R}_{b}^{\ph{b}cde}  = \mathcal{R}_{\bl{1}\,acde} \mathcal{R}_{\bl{1}\,b}^{\ph{\bl{1}\,b}cde} +
\sum_{j=2}^{n/2} \mathcal{R}_{\bl{j}\,acde} \mathcal{R}_{\bl{j}\,b}^{\ph{\bl{j}\,b}cde}\,. $$
But, if a space is maximally symmetric, any symmetric tensor of rank two built from contractions of its Riemann tensor must be proportional to its metric. Thus, a rank two symmetric tensor constructed from contractions of the curvature of the whole spacetime (\ref{Nariai3}) is a linear combination of the metrics of the 2-spaces. But, since these 2-spaces all have the same Ricci scalar,  the coefficients in front of each of the two-dimensional metrics are coincident, so that this linear combination is proportional to the metric of the whole spacetime (\ref{Nariai3}). This proves that the uncharged generalized (anti-)Nariai solution is a universal spacetime \cite{Coley-Universal,Hervik:2015mja}.

%%%
%$$ \mathcal{R}^{ab}_{\ph{ab}cd} = \frac{4\,\epsilon_1}{R_1^2}\left( - \bl{e}_1^{\;[a}\tilde{\bl{e}}_1^{\;b]}\bl{e}_{1\,[c}\tilde{\bl{e}}_{1\,d]} +
% \sum_{j=2}^{n/2}   \bl{e}_j^{\;[a}\tilde{\bl{e}}_j^{\;b]}\bl{e}_{j\,[c}\tilde{\bl{e}}_{j\,d]}  \right) . $$
%%%
%%%
%It turns out that the Minkowski and (Anti-)de Sitter spacetimes, which are maximally symmetric spacetimes, are solutions of these gravitational theories in %vacuum. Therefore, these maximally symmetric spacetimes are suitable vacuum states. In this section we will show that the generalized Nariai solutions are %also
%It turns out that
%%%

%%%%%%%%%%%%%%%%%%%%%%%%%%%%%%%%%%%%%%%%%%%%%%%%%%%%%%%%%%
%%%%%%%%%%%%%%%%%%%%%%%%%%%%%%%%%%%%%%%%%%%%%%%%%%%%%%%%%%%%%%
%%%%%%%%%%%%%%%%%%%%%%%%%%%%%%%%%%%%%%%%%%%%%%%%%%%%%%%%%%%%%%
%%%%%%%%%%%%%%%%%%%%%%%%%%%%%%%%%%%%%%%%%%%%%%%%%%%%%%%%%%%%%%
%%%%%%%%%%%%%%%%%%%%%%%%%%%%%%%%%%%%%%%%%%%%%%%%%%%%%%%%%%%%%%
%%%%%%%%%%%%%%%%%%%%%%%%%%%%%%%%%%%%%%%%%%%%%%%%%%%%%%%%%%%%%%

\section{Conclusions and Perspectives} \label{Sec. Conclusioins}

In this article we have obtained higher-dimensional generalizations of the Nariai, anti-Nariai, Bertotti-Robinson and Plaba\'{n}ski-Hacyan spacetimes that are made from the direct product of several 2-spaces of constant curvature. In $n$ dimensions, these solutions admit $n/2$ electromagnetic charges, one being electric and the rest being magnetic. These charges generate an uniform electromagnetic field throughout the spacetime. Since in four dimensions the Nariai solution can be obtained from the Schwarzschild-de Sitter spacetime in the limit that the black hole horizon and the cosmological horizon have the same temperature, we have searched for higher-dimensional black holes whose limits of thermodynamical equilibrium converge to the generalized Nariai solutions presented here.  We have also seen that similar configurations of metric and electromagnetic fields provide solutions for more general theories of gravity, the only difference being the expressions for the electromagnetic charges.

One interesting feature of the generalized Nariai solutions and their associated black holes presented here is that they possess magnetic charge, while the usual higher-dimensional generalization of the Nariai metric \cite{Kodama:2003kk,Cardoso:2004uz} and the Schwarzschild-Tangherlini black hole \cite{Tangherlini} are not magnetically charged. Therefore, it would be interesting to study the physics of the solutions obtained here. Particularly, one can seek for a rotating version of the black hole solution given by Eqs. (\ref{MetricBH}), (\ref{f}). Another nice feature of the solutions (\ref{NariMetric2}) is that the radii $R_j$ can be arbitrarily chosen. In particular, one can set $R_3$, $R_4$, $\cdots$ $R_{\frac{n-2}{2}}$ much smaller than $R_2$ and study gravitational waves and other physical phenomena in such scenario. The point being that in the latter case only four dimensions of the spacetime are accessible through low energy excitations, which is of relevance for theories and models that assume that our universe have tiny curled extra dimensions.

%%%%%%%%%%%%%%%%%%%%%%%%%%%%%%%%%%%%%%%%%%%%%%%%%%%%%%%%%%%%%%
%%%%%%%%%%%%%%%%%%%%%%%%%%%%%%%%%%%%%%%%%%%%%%%%%%%%%%%%%%%%%%
%%%%%%%%%%%%%%%%%%%%%%%%%%%%%%%%%%%%%%%%%%%%%%%%%%%%%%%%%%%%%%
%%%%%%%%%%%%%%%%%%%%%%%%%%%%%%%%%%%%%%%%%%%%%%%%%%%%%%%%%%%%%%
%%%%%%%%%%%%%%%%%%%%%%%%%%%%%%%%%%%%%%%%%%%%%%%%%%%%%%%%%%%%%%
%%%%%%%%%%%%%%%%%%%%%%%%%%%%%%%%%%%%%%%%%%%%%%%%%%%%%%%%%%%%%%

\begin{acknowledgments}
I would like to thank very much Andr\'{e}s Anabal\'{o}n for valuable suggestions in this work. I also thank Marcello Ortaggio and  Vojt\v{e}ch Pravda for pointing out some important references.
\end{acknowledgments}

\end{document}